\documentclass[11pt,aps,prb,reprint,superscriptaddress,floatfix]{revtex4-2}
\usepackage[singlespacing]{setspace}
\usepackage{graphicx}
\usepackage{amsmath}
\usepackage{amssymb}
\usepackage{mathtools}
\usepackage{xcolor}
\usepackage{braket}
\usepackage{siunitx}
\usepackage{url}
\usepackage[unicode]{hyperref}
\usepackage{booktabs}

\begin{document}

\title{Extracting the anyonic exchange phase from co-tunneling through an open quantum dot}

\author{Felix Puster}
\email{felix.puster@uni-leipzig.de}
\affiliation{Institut f\"ur Theoretische Physik, Universit\"at Leipzig, Br\"uderstrasse 16, 04103 Leipzig, Germany}

\author{Matthias Thamm}
\affiliation{Institut f\"ur Theoretische Physik, Universit\"at Leipzig, Br\"uderstrasse 16, 04103 Leipzig, Germany}

\author{Bernd Rosenow}
\affiliation{Institut f\"ur Theoretische Physik, Universit\"at Leipzig, Br\"uderstrasse 16, 04103 Leipzig, Germany}

\date{\today}

\begin{abstract}
\noindent
The elementary excitations of fractional quantum Hall states obey anyonic statistics, described by an exchange phase $\theta$ intermediate between the bosonic and fermionic values. We study an interferometer in which three chiral edge segments are pairwise connected by three quantum point contacts, so that, depending on the applied voltages, one of the segments acts as an open quantum dot: its spectrum is continuous, and tunneling through it is resonant without any gate tuning. This setup realizes the co-tunneling mechanism of a recently proposed antidot interferometer in a particularly simple setting.  Using a non-equilibrium Keldysh calculation to third order in the tunneling amplitudes, we show that the interference current contains a co-tunneling contribution in which two quasiparticles are exchanged, so that for Laughlin states the exchange phase $\theta=\pi\nu$ appears as the phase shift between the Aharonov--Bohm oscillations measured in two voltage configurations. We obtain a closed analytic expression for the interference current at finite temperature and finite separation between the quantum point contacts and show that, provided voltage-induced changes of the enclosed area are sufficiently screened, the exchange phase can be extracted from a robust phase-plateau difference. Alternatively, simultaneous measurements of the interference currents at two drains avoid this screening assumption and time-of-flight phases can be quantified analytically, allowing the exchange phase to be extracted by a controlled extrapolation with a few-percent systematic error. 
\end{abstract}

\maketitle

\section{Introduction}\label{sec:intro}
Quantum statistics in two spatial dimensions is richer than the boson-fermion dichotomy of the three-dimensional world: an exchange of two identical excitations may multiply the state by any phase factor $e^{i\theta}$, with $\theta$ interpolating between the bosonic value $0$ and the fermionic value $\pi$~\cite{Leinaas.1977,Wilczek1982PRL}. Quasiparticles with such fractional exchange statistics, anyons, are realized in the fractional quantum Hall (FQH) effect~\cite{Klitzing.1980,Tsui.1982,Laughlin.1983}. For Laughlin states at filling fraction $\nu=1/m$ with odd integer $m$, the elementary quasiparticles carry a fractional charge $e^*=\nu e$ and a fractional exchange phase $\theta=\pi\nu$~\cite{Laughlin.1983,Halperin.1984,Arovas.1984} (for reviews see Refs.~\cite{Nayak.2008,Feldman.2021,Carrega.2021}). The fractional charge was confirmed experimentally already in the 1990s, through shot noise~\cite{Kane.1994,Saminadayar.1997,dePicciotto.1997,Reznikov.1999} and through resonant tunneling across antidots~\cite{Goldman.1995}. The fractional  exchange phase, by contrast, is accessible only via controlled interference of quasiparticle trajectories and is therefore more challenging to measure.

Substantial progress has been made in recent years. Guided by the theory of Fabry--P\'erot interferometry in the FQH regime~\cite{C.Chamon.1997,Halperin.2011,Rosenow.2020,Mross.2026}, by time-domain braiding~\cite{Han.2016,Rech.2017,Lee.2019,Lee.2020,Schiller.2022,Rosenow.2025}, and by anyon-collider concepts~\cite{Rosenow.2016,Schiller.2023,Jonckheere.2023,Thamm.2024,Iyer.2024,Ronetti.2025a,Ronetti.2025b,Demazure.2026,Zhang.2025,Samal.2026}, interferometry and cross-correlation experiments in GaAs and in graphene devices have by now observed anyonic braiding~\cite{Nakamura.2020,Nakamura.2022,Nakamura.2023,Bartolomei.2020,Lee.2023,Ruelle.2023,Glidic.2023,Glidic.2024,Kundu.2023,Ruelle.2025,Ghosh.2025a,Ghosh.2025b,Samuelson.28.03.2024,Kim.2024,Kim.27.12.2024,Werkmeister.2025}. What these experiments measure is the braiding phase $2\theta$, i.e., the phase associated with a full loop of one anyon around another. Knowledge of $2\theta$, however, fixes $\theta$ only up to an additive $\pi$~\cite{Read.2024}: the values $\theta$ and $\theta+\pi$ produce the same braiding phase and cannot be told apart. Measurements with direct access to $\theta$ itself are therefore highly desirable.

Proposals for accessing the exchange phase more directly have a long history in the context of noise and current correlations~\cite{Safi.2001,Safi.2022,Kim.2005,Kim.2006,Vishveshwara.2003,Samuelsson.2004,Campagnano.2012,Campagnano.2013,Varada.2025} and of Mach--Zehnder interferometry~\cite{law2006electronic,Feldman.2007}, where $\theta$ enters two-particle interference contributions of the Hanbury Brown--Twiss (HBT) type~\cite{Brown.1954,BROWN.1956,Henny1999Science,Oliver1999Science,Buttiker1992PRB}, but typically appears in combination with non-universal parameters. In a complementary single-QPC approach, nonequilibrium fluctuation--dissipation relations have been proposed to isolate the exchange phase from noise and admittance measurements~\cite{Safi.2026}. Recently, two schemes were put forward in which the exchange phase instead appears as a  phase shift between two Aharonov--Bohm (AB) oscillatory signals measured in the same device, so that non-universal contributions largely cancel.  Kivelson and Murthy~\cite{Kivelson.2024a} suggested placing a quantum antidot~\cite{Goldman.1995,Maasilta.1998,Kataoka.1999,Goldman.2005,Averin.2007,DeLuca.2025} into one arm of a Fabry--P\'erot interferometer. On one side of the gate-tuned antidot resonance the level is empty, and quasiparticles traverse it directly; on the other side the level is occupied, and transport requires a correlated process in which the resident quasiparticle leaves before the incoming one takes its place. The two quasiparticles are thereby exchanged, and the AB signals on the two sides of the resonance differ precisely by the exchange phase. A microscopic non-equilibrium transport theory of this antidot interferometer, consistently including the fractional edge dynamics as well as the level broadening and non-equilibrium occupation of the antidot, was developed in Ref.~\cite{Thamm.2026} and showed that the bare exchange phase $\theta=\pi\nu$ can be read off from the difference between transmission-phase plateaus.  In a complementary proposal, one compares the AB oscillations of a tunneling current with those of a current cross-correlation in an HBT cross geometry interferometer. Since the former is built from single-particle and the latter from two-particle interference, their relative shift equals the exchange phase~\cite{Puster.2025}.

In this paper, instead of embedding an antidot in one arm of the interferometer, we augment the Fabry-P\'erot geometry by a single additional QPC, such that three chiral edge segments are pairwise connected by three QPCs [Fig.~\ref{fig:3QPC_setup}]. From the perspective of transport between two of the edge segments, the third segment acts as an open quantum dot: it remains connected to its own source and drain contacts, has no charging energy, and its excitation spectrum is continuous. Since the states of this open dot lie densely, tunneling through it is automatically resonant~\cite{Kane.1992,Chamon.1993}, 
and which edge segment plays the role of the dot is determined solely by the applied voltage configuration. Throughout, \emph{resonant} is meant in the sense of energy conservation into a continuum of available states with reservoir-imposed occupation, not of a gate-tuned quasi-bound level. 
As in the antidot setup, the interference current contains a co-tunneling contribution, in which a quasiparticle on the dot tunnels out towards the drain and is replaced by a quasiparticle tunneling in from the source. The two interfering amplitudes then differ by one exchange of two quasiparticles and hence by the exchange phase. The simplicity of the setup does come at a price: the charging energy of an antidot fixes the number, and thereby the charge, of the quasiparticles occupying it, providing control over the charge of the interfering quasiparticles, whereas the open dot offers no such Coulomb-blockade control. This aspect deserves attention in view of recent observations of interference dominated by coherently bunched quasiparticles of higher charge at hole-conjugate filling factors~\cite{Ghosh.2025b}. For the Laughlin states considered here, however, tunneling at a weak-backscattering QPC is dominated by quasiparticles of minimal charge $e^*=\nu e$~\cite{Kane.1992,Kane.1994}. Tunneling of larger charge $qe^*$ carries scaling dimension $q^2\nu$ and is suppressed accordingly at low energies, and interference of minimal-charge quasiparticles has been observed at $\nu=1/3$~\cite{Nakamura.2020}. The scaling dimension governing quasiparticle tunneling has recently been measured through current cross-correlations~\cite{Veillon.2024}.  The antidot and open-dot geometries are thus complementary, and the setup discussed here allows for a particularly transparent identification of the physical processes through which the exchange phase enters the transport.  A closely related three-QPC geometry was independently analyzed in a recent preprint by Sukhorukov~\cite{Sukhorukov.2026}, with emphasis on testing edge chirality and on extracting the tunneling charge and scaling dimension from the energy dependence of the interference amplitudes; here, we focus instead on the extraction of the exchange phase. 

Concretely, we compute the interference current in this three-QPC interferometer within a non-equilibrium Keldysh calculation, to third order in the tunneling amplitudes -- the lowest order at which an AB-flux-dependent contribution to the current arises. Compared with the cross geometry of Ref.~\cite{Puster.2025}, where the leading interference signals involve four tunneling amplitudes, the AB-dependent current here arises at one order lower in the tunneling and is thus parametrically larger in the weak-tunneling limit. At zero temperature, four processes contribute: two particle-like and two hole-like ones, of which one each proceeds by direct tunneling and one by co-tunneling through the open dot. Each process dominates the interference current in a distinct bias regime, so that the exchange phase can be extracted as the phase shift between AB oscillations measured in two voltage configurations. We obtain a closed analytic expression for the interference current at finite temperature and finite QPC separation in terms of generalized hypergeometric functions, which allows us to quantify the non-universal corrections to the quantized phase shift.  
At finite bias, sweeping the open-dot edge voltage from below to above the transport voltage window yields stable phase plateaus whose difference gives the exchange phase, provided voltage-induced changes of the enclosed area are sufficiently screened. Alternatively, simultaneous two-drain measurements avoid this screening requirement, although the phase difference then acquires finite time-of-flight corrections rather than forming stable plateaus. These corrections can be quantified analytically, allowing the exchange phase to be recovered by a controlled extrapolation.

The remainder of this paper is organized as follows. In Sec.~\ref{sec:model} we introduce the model of the three-QPC interferometer and set up the description of the coupled edge segments. In Sec.~\ref{sec:current} we compute the interference current to third order in the tunneling amplitudes within the Keldysh formalism. Section~\ref{sec:processes} identifies the four contributing processes and develops their interpretation as direct tunneling and as co-tunneling through an open quantum dot. In Sec.~\ref{sec:results} we present the analytic result for the interference current, the resulting phase diagram of the phase shift, and measurement protocols for extracting the exchange phase.

{This work is based on Chapter 5 of the doctoral dissertation of Felix Puster \cite{PusterDissertation.2026}.}  

\section{Model}\label{sec:model}
\begin{figure}
    \centering
    \includegraphics[width=\linewidth]{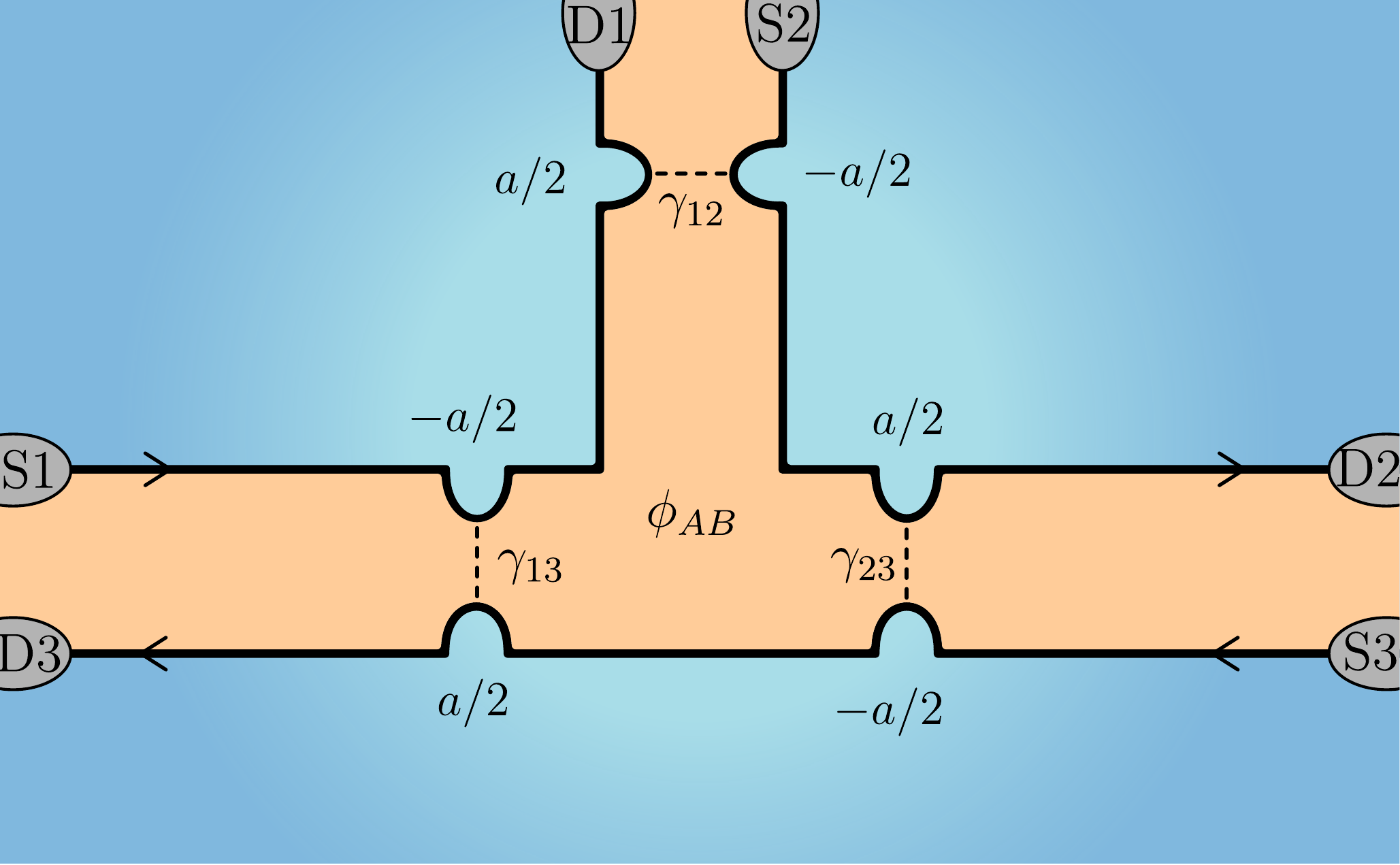}
    \caption{Interferometer formed by three chiral edge segments, each connected to its two neighbors by a QPC with tunneling amplitudes $\gamma_{ij}$. On each edge, the two QPCs are separated by a distance $a$. Voltage biases $V_1$, $V_2$, and $V_3$ are applied to sources S1, S2, and S3, respectively, and the interference current $I_{3,AB}$ 
    is measured in drain D3. The Aharonov--Bohm flux through the center area between the QPCs is denoted by $\phi_{AB}$.  Figure taken from \cite{PusterDissertation.2026}. }
    \label{fig:3QPC_setup}
\end{figure}

We consider the interferometer depicted in Fig.~\ref{fig:3QPC_setup}: three chiral edge segments of a Laughlin FQH state at filling fraction $\nu=1/m$, each connecting its own source contact to its own drain contact, are pairwise coupled by three QPCs. In the absence of tunneling, the three edge segments are independent, and each of them is described by an equilibrium chiral Luttinger liquid Hamiltonian~\cite{Wen.1990,Wen.1991}  
\begin{align}\label{eq:Heq}
    H_{0} &= \frac{\hbar v}{4\pi \nu} \sum_{i=1}^3 \int dx\, [\partial_x \phi_i(x)]^2 ,
\end{align}
where the {oscillatory part of the} chiral boson field $\phi_i$ describes quasiparticle excitations propagating with velocity $v$ along edge segment $i$. 
The equal-time commutator of the boson fields carries the anyonic statistics,
\begin{equation}\label{eq:comm}
    [\phi_i(x),\phi_j(y)]=i\theta\,\delta_{ij}\,\text{sign}(x-y) ,
\end{equation}
with the exchange phase $\theta=\pi\nu$ for Laughlin states. The operator
\begin{equation}\label{eq:qp_op}
    \psi_i(x)=(2\pi)^{-\nu/2}\,\kappa_i\, e^{i\phi_i(x)}
\end{equation}
annihilates a quasiparticle at position $x$ on edge segment $i$; the Klein factor $\kappa_i$ commutes with all boson fields and, as discussed below, carries the fractional statistics between quasiparticles residing on different segments. The quasiparticle operators~\eqref{eq:qp_op} are normalized to be dimensionless, so that the tunneling amplitudes introduced below carry units of energy, and 
the short-distance cutoff $\epsilon$ of the edge theory then enters the final results only through the combination $|\gamma_{nm}|\,(\epsilon/v)^{\nu}$ [cf.\ Eqs.~\eqref{eq:IAB_analytic} and \eqref{eq:G_fourier_app}].

Since the non-zero-mode part $H_0$ of the Hamiltonian commutes with each of the number operators, $[H_0,N_i]=0$, the voltage dependence can be removed by a gauge transformation resulting in the Hamiltonian Eq.~\eqref{eq:Heq}. As a consequence, the tunneling operators acquire explicit time- and position-dependent phase factors (for details see Appendix~\ref{app:voltageboson}).  
In this gauge, tunneling at the three QPCs is described by the time-dependent Hamiltonian
\begin{align}\label{eq:Htun}
    H_{\text{tun}}(t)&=\sum_{(n,m)} O_{nm}^+(t)\,e^{-i\frac{e^*}{\hbar}(V_n-V_m) t}  e^{i\frac{e^*}{\hbar v}(V_nx_{nm} - V_mx_{mn})} 
    \notag \\ & \hspace{1cm}
    +\text{h.c.} ,
\end{align}
with $(n,m)\in\{(1,2),(1,3),(2,3)\}$, where the tunneling operator
\begin{equation}\label{eq:tunnel_op}
    O^+_{nm}(t)=\gamma_{nm}\,\psi^\dagger_m(x_{mn},t)\,\psi^{\phantom{\dagger}}_n(x_{nm},t)
\end{equation}
moves a quasiparticle from edge segment $n$ to edge segment $m$, and its Hermitian conjugate, $O^{-}_{nm}\coloneq (O^+_{nm})^\dagger=O^+_{mn}$, moves it back. This identification implies $\gamma_{mn}=\gamma_{nm}^{*}$ for tunneling in the reversed direction. Here {$V_i$ is the voltage applied to the source contact of edge $i$ and} $x_{ij}$ is the coordinate of the tunneling point on segment $i$ that faces segment $j$. On every segment we choose coordinates such that the two tunneling points sit at $\pm a/2$, so that $a$ is the distance between the QPCs along the edge.

Consistency of the Klein-factor algebra with the commutation relations~\eqref{eq:comm} can be ensured by arranging the three segments along a single auxiliary contour, closed at infinity~\cite{Guyon.2002,Altland.2015}. Taking the segments in the order $1$, $2$, $3$ along this contour, one finds
\begin{equation}\label{eq:klein_comm}
    \kappa_i \kappa_j = \kappa_j \kappa_i\, e^{i\pi\nu\alpha_{ij}},\qquad \overline{\kappa}_i \kappa_j= \kappa_j  \overline{\kappa}_i \, e^{-i\pi\nu\alpha_{ij}},
\end{equation}
and $\kappa_i \overline{\kappa}_i=1$, where $\overline{\kappa}_i\equiv\kappa_i^\dagger$, with the antisymmetric matrix
\begin{align}\label{eq:alpha}
    \alpha &=  \left( \begin{matrix}
        0 & 1 & 1 \\
        -1 & 0 & 1  \\
        -1 & -1 & 0
    \end{matrix} \right)  .
\end{align}
Physical observables do not depend on the choice of the starting point of the contour. 
 The order of the segments along the auxiliary contour follows their downstream sequence
around the interferometer loop.

The tunneling amplitudes $\gamma_{nm}$ are complex, and the gauge-invariant combination of their phases around the interferometer loop,
\begin{equation}\label{eq:phiAB}
    \phi_{AB}=\arg\left(\gamma_{12}\,\gamma_{23}\,\gamma_{31}\right) ,
\end{equation}
contains the Aharonov--Bohm phase of the flux enclosed by the three edge segments. 
Here, the product in Eq.~\eqref{eq:phiAB} winds once around the interferometer loop in the orientation $1\to2\to3\to1$. We assume that all other contributions to the phases of the tunneling amplitudes are equal and can be absorbed into $\phi_{AB}$.  
For simplicity, we furthermore take all tunneling amplitudes to be of equal strength and distribute the AB phase symmetrically,
$\gamma_{12}=\gamma_{23}=\gamma_{31}=|\gamma|\,e^{i\phi_{AB}/3}$. 

From the tunneling operators one obtains the tunneling current operator {at $t=0$} of each QPC~\cite{C.Chamon.1997},
\begin{align}\label{eq:I_nm} 
   I_{nm} &= \frac{ie^*}{\hbar} \left[O^+_{nm}(0)\,e^{i\frac{e^*}{\hbar v}(V_nx_{nm} - V_mx_{mn})} - \text{h.c.}\right] . 
\end{align}
Current conservation implies that the current measured in drain D3 is the sum of the tunneling currents flowing into edge segment 3 at the two QPCs connecting it to segments 1 and 2,
\begin{align}\label{eq:I3} 
    &I_3  =  \langle I_{13}\rangle+\langle I_{23}\rangle\nonumber  \\
     &=  2\frac{e^*}{\hbar} \text{Re}\left[i \sum_{n\in\{1,2\}}e^{ i \frac{1}{ v} (\tilde{V}_n x_{n3}-\tilde{V}_3 x_{3n})}\,\langle O^+_{n3}(0) \rangle\right] , 
\end{align} 
where we introduced the shorthand $\tilde{V}_i= e^*V_i/\hbar$.  
 Here,  $I_3$ denotes the tunneling-induced change of the drain current (D3 also collects the Hall current $\nu e^2 V_3/h$ from $S_3$). 
Our goal is to compute the expectation value~\eqref{eq:I3} perturbatively to third order in the tunneling amplitudes. At second order, one recovers the standard tunneling current across a single QPC~\cite{Kane.1994,Chamon.1995}, which is independent of the AB phase. The leading interference contribution, which carries the dependence on $\phi_{AB}$ that we are interested in, arises at third order, where all three QPCs participate.

\section{Interference current}\label{sec:current}
Expanding the $S$ matrix on the {Keldysh} 
contour $C$ to second order in $H_{\rm tun}$, the AB-phase-dependent part of the drain current, 
$I_{3,AB}$, takes the form
\begin{align}\label{eq:IAB_keldysh} 
   I_{3,AB}  &= 2\frac{e^*}{\hbar}\,\text{Re}\Bigg[i\sum_{\substack{n,m\in\{1,2\}\\ n\neq m}} \frac{(-i)^2}{\hbar^2}\int_{C}dt_2 \int_{C}dt_3  \nonumber\\
    &  \times \left\langle T_C\, O^+_{n3}(0_{-},0_{+})\,O^-_{nm}(t_{2})\,O^-_{m3}(t_{3}) \right\rangle  
    \nonumber\\  
     &  \times e^{i(\tilde{V}_n-\tilde{V}_3)t_2-i(\tilde{V}_m-\tilde{V}_3)(t_2-t_3)}      
     \nonumber\\  
     &  \times e^{i\frac{1}{ v} \left[ x_{n3}(\tilde{V}_n+\tilde{V}_3) +x_{mn}(\tilde{V}_m+\tilde{V}_n) +x_{3m}(\tilde{V}_3+\tilde{V}_m) \right]} \Bigg] ,   
\end{align}
where $T_C$ denotes contour ordering, and we used the shorthand notation $O^+_{n3}(0_{-},0_{+})=\gamma_{n3}\,\psi_3^\dagger(x_{3n},0_{-})\,\psi_n(x_{n3},0_{+})$ for the 
vertex {at $t=0$} whose two field operators are placed on the backward ($-$) and forward ($+$) branches of the contour, respectively. We note that the same result can be obtained by iterative time-ordered perturbation theory: there, the phases generated by the Klein-factor algebra precisely reproduce the Langreth rules~\cite{Langreth.1972,Haug.2008} for the contour-ordered products, and we have verified this equivalence explicitly.

The contour-ordered expectation value in Eq.~\eqref{eq:IAB_keldysh} factorizes into products of three two-point functions of the fields $\psi_i$, multiplied by Klein-factor phases. Its evaluation simplifies considerably because transport on each edge segment is chiral: the retarded Green function $G^{\rm ret}(x,t)$ vanishes for $x<0$, and the advanced one, $G^{\rm adv}(x,t)$, for $x>0$ (up to corrections of order of the short-distance cutoff). As a consequence, time-ordered and anti-time-ordered Green functions reduce to the greater and lesser functions,
\begin{align}
    G^t(x,t)&=\begin{cases}
        G^>(x,t) & x>0\\
        G^<(x,t) & x<0
    \end{cases} ,\label{eq:G_t_simplification}\\
    G^{\bar{t}}(x,t)&=\begin{cases}
        G^<(x,t) & x>0\\
        G^>(x,t) & x<0
    \end{cases} ,\label{eq:G_t_bar_simplification}
\end{align}
so that the sums over the Keldysh branch indices of $t_2$ and $t_3$ collapse. Pairs of branch configurations combine into purely real expressions that drop out of Eq.~\eqref{eq:IAB_keldysh}, and only four terms survive (see Appendix~\ref{app:green}). These can be written compactly as
\begin{align}\label{eq:IAB_four_terms}
   &I_{3,AB} 
   = 2\frac{e^*}{\hbar^3}{\mathrm{Im}}\int_{-\infty}^{\infty}\!\!dt_2 \int_{-\infty}^{\infty}\!\!dt_3 \,\Bigg[
    e^{i\tilde{V}_1 \left(t_2-\frac{a}{v}\right)} e^{i\tilde{V}_3 \left(-t_3-\frac{a}{v}\right)} 
    \notag \\ &
    \times e^{i\tilde{V}_2 \left(t_3-t_2-\frac{a}{v}\right)} 
   \left[I_{\rm I}+I_{\rm II}\right]
    +  e^{i\tilde{V}_2 \left(t_2+\frac{a}{v}\right)} e^{i\tilde{V}_3 \left(-t_3+\frac{a}{v}\right)}  
   \notag \\ &
   \times e^{i\tilde{V}_1 \left(t_3-t_2+\frac{a}{v}\right)} 
   \left[I_{\rm III}+I_{\rm IV}\right] \Bigg], 
    \end{align}
    \begin{align}
      I_{\rm I}&=  \langle O_{21}^+(t_2)\,O_{32}^+(t_3)\,O_{13}^+(0)\rangle ,  \nonumber\\
     I_{\rm II}&=  \langle O_{32}^+(t_{3})\,O_{21}^+(t_{2})\,O_{13}^+(0)\rangle ,  \nonumber\\
     I_{\rm III}&=  \langle O_{23}^+(0)\,O_{12}^+(t_2)\,O_{31}^+(t_3)\rangle ,  \nonumber\\
     I_{\rm IV}&=  \langle O_{12}^+(t_2)\,O_{23}^+(0)\,O_{31}^+(t_3)\rangle .  \nonumber
\end{align}
The expectation values are now ordinary (fixed-order) correlation functions of the equilibrium edge theory, and the voltage phase factors uniquely determine which field operators act as creation and which as annihilation operators.

Inserting the frequency representation of the edge Green functions (Appendix~\ref{app:green}), the interference current becomes
\begin{align}\label{eq:IAB_fourier}
    I_{3,AB}&=-2\frac{e^*}{\hbar^3}|\gamma|^3\int_{-\infty}^{\infty}\!\frac{d\omega}{2\pi}\,\text{Im}\Big[e^{3i\frac{a}{v}   \omega } e^{i\phi_{AB}}\nonumber\\ &\hspace{2.3cm}\times\big(\tilde{I}_{\rm I}+\tilde{I}_{\rm II}+\tilde{I}_{\rm III}+\tilde{I}_{\rm IV}\big)\Big] ,\\
    \tilde{I}_{\rm I}&= \mathcal{G}^>(\omega-\tilde{V}_3)\,\mathcal{G}^>(\tilde{V}_1-\omega)\,\mathcal{G}^>(\omega-\tilde{V}_2) , \nonumber\\
    \tilde{I}_{\rm II}&= -e^{i\pi\nu}\, \mathcal{G}^>(\omega-\tilde{V}_3)\,\mathcal{G}^>(\tilde{V}_1-\omega)\,\mathcal{G}^>(\tilde{V}_2-\omega) , \nonumber\\
    \tilde{I}_{\rm III}&= - \mathcal{G}^>(\tilde{V}_3-\omega)\,\mathcal{G}^>(\omega-\tilde{V}_2)\,\mathcal{G}^>(\omega-\tilde{V}_1) , \nonumber\\
    \tilde{I}_{\rm IV}&= e^{i\pi\nu }\,\mathcal{G}^>(\tilde{V}_3-\omega)\,\mathcal{G}^>(\tilde{V}_2-\omega)\,\mathcal{G}^>(\omega-\tilde{V}_1) , \nonumber
\end{align} 
where $\mathcal{G}^>(\omega)$ is the Fourier transform of the greater Green function of a single edge segment. Two features of Eq.~\eqref{eq:IAB_fourier} should be noted. First, the terms $\tilde I_{\rm II}$ and $\tilde I_{\rm IV}$ carry an additional phase factor $-e^{i\pi\nu}$ relative to $\tilde I_{\rm I}$ and $\tilde I_{\rm III}$: as we discuss in the next section, these are the contributions involving co-tunneling, and the additional phase contains the anyonic exchange phase $\theta=\pi\nu$. Second, at zero temperature the greater Green function,
\begin{equation}\label{eq:G_grt_T0}
    \mathcal{G}^>(\omega)\propto \Theta(\omega)\,\omega^{\nu-1} \qquad (T=0) ,
\end{equation}
vanishes at negative frequencies, reflecting the power-law tunneling density of states of the chiral Luttinger liquid. Each of the four terms in Eq.~\eqref{eq:IAB_fourier} is therefore nonzero only in a unique frequency window, which is fixed by the applied voltages and may be empty for some voltage configurations. This observation allows us to associate each term with a distinct physical process and to selectively address individual processes by the choice of the bias voltages.

\begin{table}
\caption{Phase shift relative to the Aharonov--Bohm phase for the different processes and voltage regime in which the corresponding process is the only one contributing to $I_{3,AB}$ in the zero-temperature and zero-$a$ limit.\label{Table1}}
\begin{tabular}{ccc}
\toprule
process & phase & voltage regime\\
\midrule 
I & $0$ & $V_1>V_3>V_2$\\ 
II & $\pi(\nu-1)$ & $V_2>V_1>V_3$\\ 
III & $\pi$ & $V_3>V_1>V_2$ \\ 
IV & $\pi\nu$ & $V_2>V_3>V_1$\\
\bottomrule
\end{tabular}
\end{table}

\section{Interference processes}\label{sec:processes}

\begin{figure}
    \centering
    \includegraphics[width=1\linewidth]{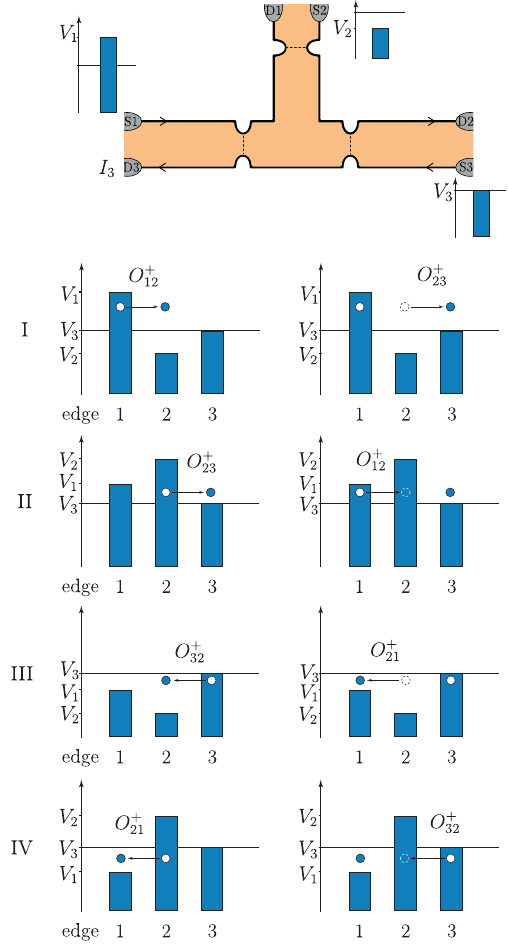}
    \caption{Schematic voltage configurations relevant for the four processes. The top panel depicts the setup and voltage definitions, where voltage $V_3$ on edge 3 is used as the reference. The other panels show the tunneling paths for processes I--IV, which  interfere with the reference process of direct tunneling between edges 1 and 3 to give rise to the current $I_{3, AB}$. An operator $O_{ij}^+$ describes tunneling of a particle (blue circle) from edge $i$ to edge $j$, which may leave behind a hole (white circle). For the co-tunneling processes in II and IV, first tunneling of a particle creates a hole, which is then filled by a second particle.  Processes I and II differ by an exchange of operators and have reference process $O^+_{13}\ket{\psi_0}$; III and IV differ by an exchange of operators and have reference process $O^+_{31}\ket{\psi_0}$. Here, $\ket{\psi_0}$ denotes the initial many-body state.}
    \label{fig:processes}
\end{figure}

\begin{figure}
    \centering
    \includegraphics[width=0.85\linewidth]{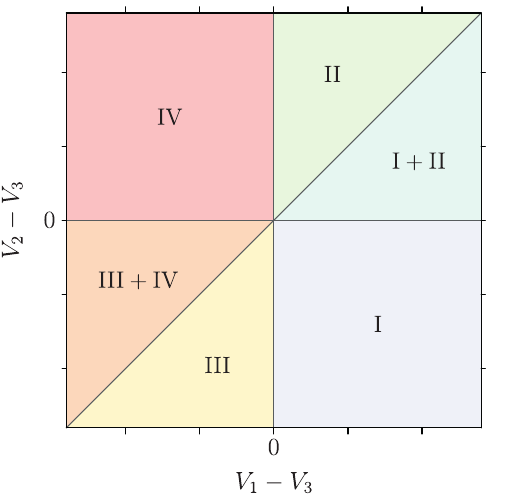}
    \caption{Region diagram for the processes at zero temperature as a function of the voltages $V_1-V_3$ and $V_2-V_3$ using $V_3$ as a reference voltage. The colors show the different regions and the labels indicate the process corresponding to the region. In the lower half of the upper right quadrant and in the upper half of the lower left quadrant, a mixture of two processes occurs. 
    \label{fig:phasediagram}}
\end{figure}

The four terms of Eqs.~\eqref{eq:IAB_four_terms} and \eqref{eq:IAB_fourier} are in one-to-one correspondence with four interference processes, which we illustrate in Fig.~\ref{fig:processes}. In each panel, we show the path that interferes with the reference process of tunneling directly between edges 1 and 3.  
Terms $I_{\rm I}$ and $I_{\rm II}$ 
describe particle-like processes, in which a net quasiparticle current flows into edge segment 3, while terms $I_{\rm III}$ and $I_{\rm IV}$
describe the corresponding hole-like processes with the current direction reversed. Within each pair, one process,  
terms $I_{\rm I}$ and $I_{\rm III}$,
proceeds by direct tunneling only, while the other,
terms $I_{\rm II}$ and $I_{\rm IV}$, involves co-tunneling and carries the additional factor $-e^{i\pi\nu}$ in Eq.~\eqref{eq:IAB_fourier}. The identification is confirmed by the frequency windows in which the four terms contribute at zero temperature: from Eqs.~\eqref{eq:IAB_fourier} and \eqref{eq:G_grt_T0} one finds that term $\tilde I_{\rm I}$ is supported on $\max(\tilde V_3,\tilde V_2)<\omega<\tilde V_1$, term $\tilde I_{\rm II}$ on $\tilde V_3<\omega<\min(\tilde V_1,\tilde V_2)$, term $\tilde I_{\rm III}$ on $\max(\tilde V_1,\tilde V_2)<\omega<\tilde V_3$, and term $\tilde I_{\rm IV}$ on $\tilde V_1<\omega<\min(\tilde V_3,\tilde V_2)$. Each term therefore requires a distinct configuration of the applied voltages. In particular, for $V_1>  V_3>V_2$ only the process I contributes, and for $V_3<V_1<V_2$ only process II contributes at zero temperature. The processes, voltage windows, and phase contributions are summarized in Table~\ref{Table1} and a phase diagram is shown in Fig.~\ref{fig:phasediagram}.

The process I is a single-particle interference process: a quasiparticle incoming on edge segment 1 can tunnel into segment 3 in two distinct ways, either directly at the QPC connecting segments 1 and 3, or via segment 2, by tunneling at the two QPCs that connect segment 1 to segment 2 and segment 2 to segment 3. The two amplitudes enclose the interferometer loop, and their interference depends on the bare AB phase $\phi_{AB}$. This process requires the states at the quasiparticle's energy to be unoccupied on segments 2 and 3, which is ensured in the voltage window given above.

If, instead, a state at this energy on segment 2 is already occupied, a two-particle interference process can contribute to the current in D3. This process, shown in panel II, can be understood in two equivalent ways. In the first view, it is a two-particle interference process: two quasiparticles are incoming on segments 1 and 2, and there are two amplitudes in which one of them ends up on segment 1 and the other on segment 3. Since the quasiparticles are indistinguishable, the two amplitudes interfere. Because the corresponding outgoing states differ by one exchange of the two quasiparticles, the interference encodes the fractional exchange phase $\theta=\pi\nu$. { The Klein factors for $O_{21}^+ O_{32}^+$ are $(\bar{\kappa}_1\kappa_2)(\bar{\kappa}_2\kappa_3) = (\bar{\kappa}_2\kappa_3)(\bar{\kappa}_1\kappa_2) e^{-i\pi\nu\alpha_{21}} e^{i\pi\nu\alpha_{23}}e^{-i\pi\nu\alpha_{13}} = (\bar{\kappa}_2\kappa_3)(\bar{\kappa}_1\kappa_2) e^{i\pi\nu}$, such that $O_{21}^+ O_{32}^+ O_{13}^+$ and  $O_{32}^+O_{21}^+  O_{13}^+$ differ by a Klein factor exchange phase $\pi\nu$.} 

In the second view, edge segment 1 acts as an open quantum dot. The quasiparticle incoming on segment 2 cannot tunnel through the dot into segment 3, since the dot state at its energy is occupied. There is, however, a cooperative tunneling process: the quasiparticle on the dot tunnels out into segment 3 and is subsequently replaced by the quasiparticle from segment 2, which tunnels into the hole left behind by the first quasiparticle, after this hole has propagated to the QPC connecting segments 1 and 2. In this co-tunneling process the two quasiparticles are exchanged, so that the exchange phase enters the interference with the direct amplitude, in which the quasiparticle from segment 2 tunnels into segment 3 at the QPC connecting these two segments. This second view is in the spirit of the antidot interferometer proposed in Ref.~\cite{Kivelson.2024a} and analyzed microscopically in Ref.~\cite{Thamm.2026}, with the Coulomb-blockaded antidot replaced by the open dot formed by edge segment 1. We note that the source and drain contacts of the dot segment do not dephase this process: on every segment the two tunneling points lie between its source and its drain, so the propagating hole never enters a reservoir, and the openness of the dot manifests itself only through its continuous spectrum and its unquantized occupation.

As reflected by the factor $-e^{i\pi\nu}$ of the term $\tilde I_{\rm II}$ in Eq.~\eqref{eq:IAB_fourier}, the co-tunneling process carries a phase shift of $\pi$ in addition to the exchange phase. This shift can be understood as arising from the sign of the energy of the virtual intermediate state relative to the initial and final states of the tunneling process~\cite{Casper.2018}. In the antidot geometry, where the dot has well-separated levels, this $\pi$ is compensated away from resonance by the sign of the Breit-Wigner amplitude of the direct transmission channel~\cite{Kivelson.2024a,Thamm.2026,Casper.2018}. For the open dot considered here there is no Breit-Wigner resonance, and the direct and co-tunneling processes differ by the total phase shift $\pi\nu-\pi$.

{For the process in panel IV,} edge segment 2 acts as the open quantum dot, and a quasiparticle from segment 3 co-tunnels through segment 2 into segment 1, i.e., the current at D3 flows outward. The minus sign of this hole-like current compensates the co-tunneling minus sign. In Eq.~\eqref{eq:IAB_fourier}, the term $\tilde I_{\rm IV}$ carries the factor $+e^{i\pi\nu}$ relative to $\tilde I_{\rm I}$.   
Which edge segment plays the role of the open dot is thus not fixed by the geometry but selected by the voltage configuration.

We close this section by noting why the exchange phase appears in this three-QPC geometry but not in a Fabry-P\'erot interferometer. In the latter, the leading interference contribution arises at second order in the tunneling, the two field operators of each tunneling operator act at equal times, and the Klein factors drop out of the result~\cite{C.Chamon.1997}. In the present geometry, the co-tunneling contributions contain a contraction in which a tunneling operator acting at an intermediate time is exchanged with a field operator of a different edge segment. It is precisely this exchange, mediated by the Klein-factor algebra of Eq.~\eqref{eq:klein_comm}, that imprints the statistical phase $e^{i\pi\nu}$ onto the interference current. As a further consistency check on this mechanism, Appendix~\ref{app:nu1} shows that for $\nu=1$ the four terms of Eq.~\eqref{eq:IAB_fourier} collapse into the Landauer--B\"uttiker expression for free chiral fermions scattering at three point contacts.

\section{Results}\label{sec:results}
At zero temperature, the four processes identified in the previous section contribute in sharply defined voltage windows, and the associated phase shifts take the quantized values discussed there. At finite temperature, the step functions in Eq.~\eqref{eq:G_grt_T0} are smeared, all four processes contribute for any choice of the bias voltages, and the phase shift of the interference current evolves continuously between the quantized values. 
Within the approximation $1-2k_BT\epsilon/\hbar v\approx1$, which is excellent for the small short-distance cutoff $\epsilon$, the frequency integral in Eq.~\eqref{eq:IAB_fourier} can be evaluated in closed form at finite temperature by summing residues (see Appendix~\ref{app:residues}). We obtain
\begin{widetext}
    \begin{align}\label{eq:IAB_analytic} 
        I_{3,AB} (V_1,V_2,V_3)=&-2\frac{e^*\epsilon^{3\nu}\tilde{T}^{3 \nu -2}}{v^{3\nu}\hbar^3}|\gamma|^3 \frac{(2\pi)^2}{\Gamma (\nu )^2}\,\text{Im}\Bigg\{  e^{-3 \pi \nu \tilde{a} \tilde{T}+i \phi _{\text{AB}}+\frac{5 i \pi \nu}{2}+\frac{\tilde{V}_1+\tilde{V}_2-2\tilde{V}_3}{2 \tilde{T}}} \nonumber\\
        &\times \left[\frac{e^{3 i \tilde{a} \tilde{V}_1 +\frac{\tilde{V}_1-\tilde{V}_3}{2 \tilde{T}}} \,
_3\!\tilde{F}_2\left(\nu ,\nu -\frac{i (\tilde{V}_1-\tilde{V}_3)}{2 \pi  \tilde{T}},\nu-i\frac{\tilde{V}_1- \tilde{V}_2}{2 \pi  \tilde{T}};1-i\frac{\tilde{V}_1-\tilde{V}_3}{2 \pi  \tilde{T}},1-i\frac{  \tilde{V}_1- \tilde{V}_2}{2 \pi  \tilde{T}};e^{-6 \tilde{a} \pi
\tilde{T}}\right)}{\left(e^{\frac{\tilde{V}_1-\tilde{V}_3}{\tilde{T}}+2 i \pi  \nu }-1\right) \left(e^{\frac{\tilde{V}_1-\tilde{V}_3}{\tilde{T}}+2 i \pi  \nu }-e^{\frac{\tilde{V}_2-\tilde{V}_3}{\tilde{T}}}\right) \Gamma \left(1-\nu +i\frac{ \tilde{V}_1 -\tilde{V}_3}{2 \pi  \tilde{T}}\right)
\Gamma \left(1-\nu +i\frac{ \tilde{V}_1-\tilde{V}_2}{2 \pi  \tilde{T}}\right)}\right.\nonumber\\
& \quad \quad +\left.\frac{e^{3i \tilde{a} \tilde{V}_3} \, _3\!\tilde{F}_2\left(\nu ,\nu +i\frac{  \tilde{V}_1-\tilde{V}_3}{2 \pi  \tilde{T}},\nu + i\frac{ \tilde{V}_2-\tilde{V}_3}{2
\pi  \tilde{T}};i\frac{\tilde{V}_1-\tilde{V}_3}{2 \pi  \tilde{T}}+1,i\frac{ \tilde{V}_2-\tilde{V}_3}{2 \pi  \tilde{T}}+1;e^{-6 \tilde{a} \pi  \tilde{T}}\right)}{\left(-e^{\frac{\tilde{V}_1-\tilde{V}_3}{\tilde{T}}}+e^{2 i \pi  \nu }\right) \left(e^{\frac{\tilde{V}_2-\tilde{V}_3}{\tilde{T}}}-e^{2 i \pi  \nu
}\right) \Gamma \left(1-\nu -i\frac{ \tilde{V}_1-\tilde{V}_3}{2 \pi  \tilde{T}}\right) \Gamma \left(1-\nu - i\frac{ \tilde{V}_2-\tilde{V}_3}{2 \pi  \tilde{T}}\right)}\right]\Bigg\} ,
    \end{align} 
\end{widetext}
where $_3\tilde{F}_2$ denotes the regularized generalized hypergeometric function, and we abbreviated $\tilde{T}=k_B T/\hbar$, $\tilde{V}_i = e^*V_i/\hbar$, and $\tilde{a}=a/v$. We have validated Eq.~\eqref{eq:IAB_analytic} against direct numerical integration of Eq.~\eqref{eq:IAB_fourier}.  
We now write the interference current as
\begin{equation}\label{eq:define_delta}
    I_{3,AB} (V_1,V_2,V_3)=- \tilde{I}(V_1,V_2,V_3)\sin(\phi_{AB}+\delta(V_1,V_2,V_3)) ,
\end{equation}
which defines the phase shift $\delta(V_1,V_2,V_3)$ of its AB oscillations. Here, $\tilde{I}(V_1,V_2,V_3)$ is independent of the Aharonov--Bohm flux.

\subsection{Extraction of the exchange phase}

 We now describe a voltage-sweep protocol that directly connects the direct tunneling  and co-tunneling phase plateaus. We apply a symmetric bias between edges 1 and 3, $V_1=-V_3>0$, and vary the voltage $V_2$ of the remaining edge. This choice keeps the transport window between edges 1 and 3 centered at zero voltage while continuously changing the occupation of edge 2. We assume sufficient electrostatic screening that this voltage sweep does not appreciably change the enclosed area~\cite{Halperin.2011,Rosenow.2007}. The corresponding electrostatic lever arm is device specific, and its required magnitude can be estimated: for the readout points $V_2=\pm\SI{80}{\micro\volt}$, keeping this contribution below $10\%$ of $\pi/3$ requires $|\phi_{AB}(+\SI{80}{\micro\volt})-\phi_{AB}(-\SI{80}{\micro\volt})|<\pi/30$, corresponding to an average slope below $6.5\times10^{-4}\,\mathrm{rad}/\si{\micro\volt}$. A residual electrostatic contribution can be diagnosed as a smooth drift of the nominal phase plateaus and included as a background in the phase fit. The simultaneous two-drain protocol of Sec.~\ref{sec:simultaneous} avoids this screening assumption.

\begin{figure}
    \centering
    \vspace{0.2cm}
    \includegraphics[width=0.85\linewidth]{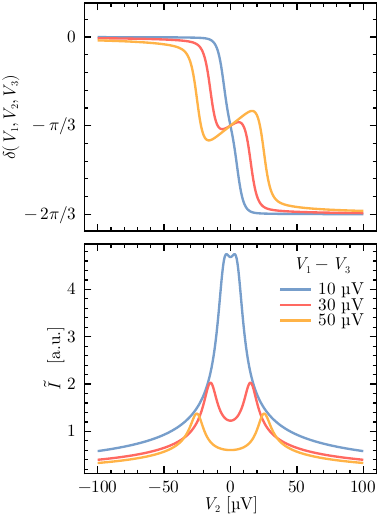}
    \caption{Phase shift $\delta$ and amplitude $\tilde{I}$ of the Aharonov--Bohm dependent part of the tunneling current in drain D3, Eq.~\eqref{eq:define_delta}, as a function of $V_2$ for different values of the voltage bias $V_1-V_3$ with $V_1=-V_3 >0$. Results are shown for a temperature of $\SI{10}{\milli\kelvin}$, edge velocity $v=10^5\,\rm m/s$, and a QPC distance of $a=\SI{1}{\micro\meter}$. For $V_2 \ll V_3$, only process I contributes and the phase shift to the Aharonov--Bohm phase shows a stable plateau at $0$. For $V_2\gg V_1$, only process  II contributes such that $\delta$ plateaus at $\pi(1/3-1)$. In between the phase evolves non-monotonically. Due to the density of states anomaly, the amplitude is peaked at the Fermi levels of edges 1 and 3, $V_2 = \pm(V_1-V_3)/2$. } 
    \label{fig:V2_phase}
\end{figure}

{At zero temperature, only processes I and II can contribute for $V_1>V_3$: process I is supported when $V_2<V_1$, whereas process II is supported when $V_2>V_3$. The sweep of $V_2$ therefore passes successively through a regime dominated by the direct single-particle process I, a mixed regime in which processes I and II coexist, and a regime dominated by the particle-like co-tunneling process II. For $V_2$ sufficiently far below $V_3$, edge 2 is empty throughout the transport window. The co-tunneling contribution is then exponentially suppressed at finite temperature, and the interference current is dominated by process I. Its Aharonov--Bohm oscillations carry no statistical phase shift. Conversely, for $V_2$ sufficiently far above $V_1$, edge 2 is occupied throughout the transport window and process II dominates. The additional factor $-e^{i\pi\nu}$ of this process produces the asymptotic phase $\pi(\nu-1)$. 
At finite temperature and finite QPC separation, the plateau values receive small non-universal corrections, but the two limiting regimes remain continuously connected to the quantized values $0$ and $\pi(\nu-1)$. In the intermediate interval $V_3<V_2<V_1$,  both particle-like processes contribute. At zero temperature, their frequency windows partition the transport window:  $\tilde V_2<\omega<\tilde V_1$  for process I and $\tilde V_3<\omega<\tilde V_2$  for process II. Increasing $V_2$ therefore continuously transfers spectral weight from the direct process to the co-tunneling process. Since the two contributions carry different statistical phases and are averaged with the energy-dependent propagation factor $\exp(3ia\omega/v)$, the phase of their coherent sum  displays the non-monotonic evolution visible in the upper panel of Fig.~\ref{fig:V2_phase}. The crossover is particularly pronounced when $V_2$ passes either boundary of the transport window. At $V_2=V_3$, the frequency window of process II opens, whereas at $V_2=V_1$ the window of process I closes. At these two thresholds, singularities of the chiral-Luttinger-liquid spectral functions approach one another. For $\nu=1/3$, the power-law tunneling density of states, $\mathcal G^>(\omega)\propto\omega^{-2/3}$ at zero temperature, therefore strongly enhances the interference amplitude. Finite temperature rounds the singularities into the two peaks shown in the lower panel of Fig.~\ref{fig:V2_phase}. }

\begin{figure}
    \centering
    \vspace{0.2cm}
    \includegraphics[width=0.85\linewidth]{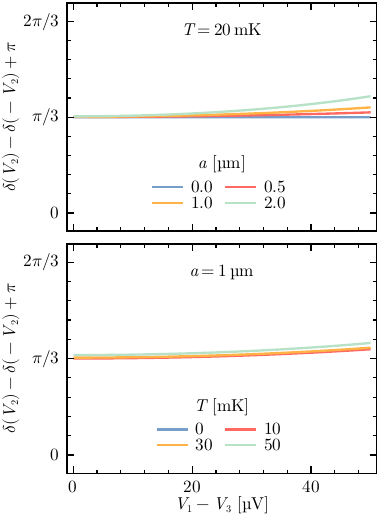}
    \caption{Exchange phase $\delta(V_1,V,V_3)-\delta(V_1,-V,V_3)+\pi$ extracted from 
     plateau values obtained as in Fig.~\ref{fig:V2_phase} at $V=\SI{80}{\micro\volt}$ as a function of the bias $V_1-V_3$. The upper panel shows results for a temperature of $T=\SI{20}{\milli\kelvin}$ for different QPC distances $a$. The lower panel uses $a=\SI{1}{\micro\meter}$ and shows extracted exchange phases for different temperatures.  We find that for a wide range of biases, temperature, and QPC distances, the exchange phase of $\pi/3$ can reliably be extracted.}
    \label{fig:error}
\end{figure}
 
Figure~\ref{fig:V2_phase} illustrates this sweep for $\nu=1/3$, $T=\SI{10}{\milli\kelvin}$, $v=10^5\,\mathrm{m/s}$, and $a=\SI{1}{\micro\meter}$. For all bias values shown, the phase approaches the single-particle plateau near zero for $V_2<V_3$ and the co-tunneling plateau near $\pi(\nu-1)=-2\pi/3$ for $V_2>V_1$. The plateau separation therefore contains the exchange phase, apart from the additional $\pi$ associated with the sign of the co-tunneling amplitude. To quantify the extraction, we evaluate the two plateaus at the fixed voltages $V_2^-=-V$ and $V_2^+=+V$ with $V=\SI{80}{\micro\volt}$,  and  estimate the exchange phase from $\delta\left( V_1, V, V_3\right) - \delta(V_1,-V,V_3) + \pi$.   
 The added $\pi$ compensates a non-statistical sign fixed by the
energy denominator of the co-tunneling amplitude. Since it is independent of  $\nu$, the $\theta$  versus $\theta + \pi$ ambiguity of braiding measurements is not reintroduced. 
 Figure~\ref{fig:error} shows the extracted phase as a function of the bias $V_1-V_3$. The upper panel fixes $T=\SI{20}{\milli\kelvin}$ and compares different QPC separations, whereas the lower panel fixes $a=\SI{1}{\micro\meter}$ and compares different temperatures. Two independent effects determine the deviation from $\pi\nu$. First, finite temperature broadens the thresholds at $V_2=V_3$ and $V_2=V_1$, so that thermally activated contributions of the subdominant process extend onto the nominal plateaus. Second, a finite QPC separation produces the energy-dependent phase $3a\omega/v$, whose variation over the transport window grows with both $a/v$ and the applied bias.  At finite temperature, a well-resolved plateau additionally requires $e^* \left( V - {|V_1-V_3|}/{2} \right) \gg k_BT$. If $|V_1-V_3|$ approaches $2V$, the two readout points approach the Fermi levels of edges 1 and 3 and no longer probe the asymptotic plateaus. At smaller bias, thermal mixing becomes more important relative to the width of the transport window, whereas at larger bias or larger $a$ the dynamical propagation phase produces stronger corrections. The competition between these effects leaves a broad intermediate regime in which the extracted value is close to the ideal exchange phase $\pi/3$.

In the following, we discuss the range of validity of the perturbative expansion  \cite{Puster.2025,Thamm.2026}.
Quasiparticle tunneling at a QPC is a relevant perturbation for $\nu<1$~\cite{Kane.1992}, so that the expansion is controlled by the dimensionless coupling at the infrared scale $E$ of the transport,
\begin{equation}\label{eq:g_coupling}
    g(E)=\frac{|\gamma|\,\epsilon^{\nu}}{(\hbar v)^{\nu}}\,E^{\nu-1} ,
\end{equation}
in terms of which the prefactor of the interference current \eqref{eq:IAB_analytic} takes the form 
$I_{3,AB} \propto (e^*k_BT/\hbar)\,g(k_BT)^{3}$. All results presented here require $g(E_{ij})\ll1$ at 
every scale $E_{ij}=\max(k_BT,e^*|V_i-V_j|)$ for which a QPC connects edges $i$ and $j$.
{For the plateau extraction at
$V_1-V_3=\SI{50}{\micro\volt}$ in the upper panel of
Fig.~\ref{fig:error}, where
$V_1=-V_3=\SI{25}{\micro\volt}$,
$V_2=\pm\SI{80}{\micro\volt}$, and
$T=\SI{20}{\milli\kelvin}$,
the smallest bias-induced infrared cutoff is $E_{13}/k_B\approx \SI{193}{\milli\kelvin}$  for $\nu=1/3$. This is approximately ten times the temperature, and all three QPCs are cut off by the applied biases rather
than by temperature at this operating point, so the expansion is not pushed toward the infrared strong-backscattering regime.}
An AB-dependent contribution requires the oriented loop product $\gamma_{12}\gamma_{23}\gamma_{31}$ or its conjugate, i.e., three unpaired tunneling events. Quasiparticle-number conservation on each edge segment forces any further tunneling operators to occur in conjugate pairs, so the leading corrections to the interference signal arise at fifth order and are suppressed by a relative factor $g^2$. Since the backscattering probability of a single QPC at the energy scale $E$ is of order $g(E)^2$, a QPC operated at a few-percent reflection at the scale $e^*V$ of the protocols below corresponds to $g\approx0.2$, for which the relative corrections $\sim g^2$ are at the percent level.
 Near the thresholds $V_2 = V_3$ and $V_2 = V_1$,
where the interference amplitude is enhanced, two singularities of the integrand of Eq.~\eqref{eq:IAB_fourier} merge. Within strips $|V_2 - V_{1,3}| \lesssim k_BT /e^*$ the expansion is then controlled by temperature alone. The readout
points $V_2 = \pm 80 \mu V$ lie far outside these strips.

Finally, we estimate the magnitude of the interference signal and comment on decoherence beyond finite temperature. At the operating point, the current scale set by Eq.~\eqref{eq:IAB_analytic} is $I_{3,AB} 
\sim g^3\,(e^*)^2V/\hbar\approx10\,$pA for $g\approx0.2$, which we expect to be in range of standard lock-in techniques.  
The only decoherence contained in our treatment is thermal. Slow environmental noise, most prominently telegraph noise from individual bulk quasiparticle rearrangements~\cite{Werkmeister.2025} and gate-induced fluctuations of the enclosed area, reduces the visibility of the AB oscillations, but affects the extracted exchange phase only through drifts between the two sequential measurements. The most important source of errors is due to 
 comparing Aharonov--Bohm phases for values of  $V_2$ well below and well above the transport window.  Since this large voltage excursion can modify the edge charge and thereby the enclosed area, identifying the plateau difference with the exchange phase relies on essentially perfect electrostatic screening, or at least careful subtraction of the change in AB phase due to the change in $V_2$.

\subsection{Simultaneous measurement of two interference currents}\label{sec:simultaneous}

As an alternative to the sequential measurement of the two interference signals, for a given  $V>0$ one can use the balanced configuration $V_1=V/2$, $V_2=-V/2$,  $V_3=0$,  and measure the interference currents at drains D3 and D2 simultaneously.
In this way, one avoids the requirements on electrostatic screening of voltage changes discussed above.
For a finite propagation time between the QPCs, however, the phase difference between interference currents measured at
D3 and D2 contains bias-dependent time-of-flight contributions instead of forming extended plateaus. We show below that these contributions are sufficiently controlled that a linear extrapolation of the voltage-dominated branch to zero bias recovers the exchange phase with only a few percent systematic error. 

 The interference current at drain D2 follows from our result for drain D3 by relabeling the edge segments counterclockwise,
$I_{2,AB}(V_1,V_2,V_3) = I_{3,AB}(V_3,V_1,V_2)$.  Relative to edge segment 2, segment 3 is then at a positive bias $V/2$, while segment 1 is at the even larger bias $V$. The AB-dependent part of the current in D3 is therefore dominated by process I with phase $0$, while the current in D2 is dominated by process II. Thus the currents have a Klein factor related relative phase shift of $\pi(\nu-1)$. 

\begin{figure}
    \centering
    \vspace{0.2cm}
    \includegraphics[width=0.85\linewidth]{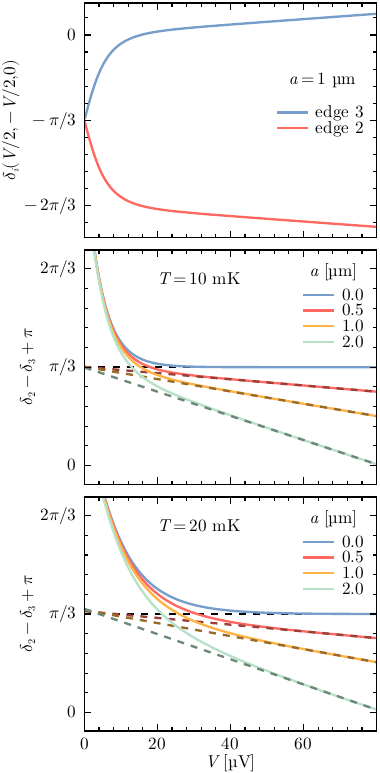}
    \caption{Estimation of the exchange phase from simultaneous measurements of the currents in drains D3 and D2 for $V_1=-V_2\equiv V/2$ and $V_3=0$ at temperature $T=\SI{10}{\milli\kelvin}$ (upper and center panels) and  $T=\SI{20}{\milli\kelvin}$ (lower panel). The upper panel shows the phase shift to the Aharonov--Bohm phase of the current on edges 2 ($\delta_2$, red) and 3 ($\delta_3$, blue) for QPC distance $a=\SI{1}{\micro\meter}$. The center and lower panels depict the estimated exchange phase from the difference of these phases and an additional $\pi$ shift for various values of $a$. Because both currents are measured at the same voltages, common electrostatic phase shifts cancel, while finite-temperature and time-of-flight corrections produce a voltage dependence rather than stable plateaus. Linear fits to the region $V\in[\SI{60}{\micro\volt}, \SI{80}{\micro\volt}]$
    are shown as dashed lines. From the $V=0$ intercepts of these fits, the exchange phase can be obtained with relative errors well below $10\%$.}
    \label{fig:simult}
\end{figure}

 Fig.~\ref{fig:simult} depicts the phase shift $\delta_i$ of the current in drain $i$ as a function of the bias $V$ (upper panel). We find that for small $V$, the currents are in a temperature-dominated regime, where the mix of several processes results in phase shifts approaching each other. When approaching the voltage-dominated regime, the phase difference passes through its ideal value $\pi(\nu-1)$ before dynamical phase corrections result in a linear voltage dependence of the phases (lower panel). The magnitude of the slope of these corrections increases with increasing QPC distance $a$, and stable plateaus are only reached in the $a\to0$ limit (center panel for $T=\SI{10}{\milli\kelvin}$ and lower panel for $T=\SI{20}{\milli\kelvin}$, blue).  

We can also quantify the finite-temperature time-of-flight phase correction $\delta_{\rm tof}(V)$ analytically (see Appendix~\ref{app:tof_simultaneous}). For the parametrization $V_1=-V_2=V/2$ used here, the phase difference obeys $\delta_2(V)-\delta_3(V)+\pi=\pi\nu+2\delta_{\rm tof}(V)$ modulo $2\pi$. In the bias-dominated, short-time-of-flight regime $e^*V/(2k_BT)\gg1$ and $3ae^*V/(2v\hbar)\ll1$, we find
\begin{align}
&\delta_{\rm tof}(V)
\approx 
\sin(\pi\nu) \exp\!\left(-\frac{e^*V}{2k_BT}\right)
\notag\\ &
-\frac{3}{2}
\frac{\mathrm{B}\!\left(\frac{\nu+1}{2},\nu\right)}
     {\mathrm{B}\!\left(\frac{\nu}{2},\nu\right)}
\frac{ae^*V}{v\hbar}
\left[
1-2\pi^2\nu(1-\nu)
\left(\frac{k_BT}{e^*V}\right)^2
\right] ,
\end{align}
where $\mathrm{B}(x,y)={\Gamma(x)\Gamma(y)}/{\Gamma(x+y)}$ is the Beta function and the approximation is up to terms of order $\mathcal{O}\!\Big[ e^{-e^*V/k_BT}, e^{-e^*V/(2k_BT)}\frac{ae^*V}{v\hbar}, \big(\frac{ae^*V}{v\hbar}\big)^3, \frac{ae^*V}{v\hbar} \big(\frac{k_BT}{e^*V}\big)^4 \Big]$. In the zero temperature limit, the expansion is given by 
\begin{align}
&\delta_{\rm tof}(V) \xrightarrow{T\to0}{} 
-\frac{3}{2}
\frac{\mathrm{B}\!\left(\frac{\nu+1}{2},\nu\right)}
     {\mathrm{B}\!\left(\frac{\nu}{2},\nu\right)}
\frac{ae^*V}{v\hbar}
+
\mathcal{O}\!\left[
\left(\frac{ae^*V}{v\hbar}\right)^3
\right].
\end{align}
Consequently, the voltage-dominated branch of $\delta_2(V)-\delta_3(V)+\pi$ is approximately linear, with an intercept $\pi\nu$ and a slope proportional to $a/v$. The exchange phase can therefore be obtained from the zero-voltage intercept of a linear fit to this branch (dashed lines in center and lower panel). In addition, a comparison of the fitted slope to the analytical value can serve as a test that the correct regime is probed, in analogy to the extraction of the edge-state velocity from the bias dependence of Fabry--P\'erot interference fringes~\cite{C.Chamon.1997,McClure.2009}.

{For $\nu=1/3$, $e^*=e/3$, $v=10^5\,\mathrm{m/s}$, and $a=0.5$, $1$, and $\SI{2}{\micro\meter}$, respectively, a  linear fit to the region $V\in[\SI{60}{\micro\volt},\,\SI{80}{\micro\volt}]$ 
gives relative deviations of the extrapolated intercept from $\pi/3$ of $+0.27\%$, $+0.41\%$, and $-0.29\%$ at $T=10\,\mathrm{mK}$, and $+2.97\%$, $+4.12\%$, and $+5.25\%$ at $T=20\,\mathrm{mK}$. 
Thus, the exchange phase can be extracted with corrections well below ten percent from the simultaneous measurement protocol.}

\section{Summary}\label{sec:summary}
In this paper, we have shown that an interferometer built from three chiral edge segments, pairwise connected by three QPCs, enables two-particle interference processes that encode the fractional exchange phase of the interfering anyons. Such processes are absent in a Fabry-P\'erot interferometer. Here, they arise because one edge segment acts as an open quantum dot, through which quasiparticles can co-tunnel, and which edge segment plays this role is selected by the applied voltage configuration alone. The particle-like co-tunneling process carries a total phase shift of $\pi\nu-\pi$. 

From a non-equilibrium Keldysh calculation, we obtained a closed analytic expression for the interference current to leading order in the tunneling, valid at finite temperature and finite QPC separation. 
Our analytic result shows that the exchange phase can be extracted from the difference between the phase plateaus obtained by sweeping $V_2$ from below $V_3$ to above $V_1$ at fixed symmetric bias $V_1=-V_3>0$, provided voltage-induced changes of the enclosed area are sufficiently screened. This screening requirement is avoided by simultaneously measuring the interference currents at two drains at the same voltage configuration. Although their phase difference then acquires time-of-flight corrections rather than forming stable plateaus, these corrections can be  controlled  in the low temperature regime, and the exchange phase can be obtained from the zero-voltage intercept of a linear fit to the voltage-dominated branch. For $\nu=1/3$, $e^*=e/3$, $v=10^5\,\mathrm{m/s}$, and $a=0.5$, $1$, and $\SI{2}{\micro\meter}$, extracted phases have relative deviations from $\pi/3$ of $-0.29$--$0.41\%$ at $T=\SI{10}{\milli\kelvin}$  and $2.97$--$5.25\%$ at $T=\SI{20}{\milli\kelvin}$. 

Our results were derived for Laughlin states with a single chiral edge mode, and the filling fraction $\nu$ plays two logically distinct roles in them. The statistical factor $e^{i\pi\nu}$ of the co-tunneling terms is topological: it originates from the Klein-factor algebra and is insensitive to microscopic details of the edge. The exponents $\nu-1$ of the tunneling density of states, by contrast, can be renormalized by edge reconstruction or coupling to additional modes. The quantized plateau values $0$, $\pi\nu$, $\pi\nu-\pi$, and $\pi$ of the phase shift are fixed by the statistical factors alone and therefore survive such a renormalization;  
hole-conjugate and multi-mode edges lie outside the present treatment. 

Given ongoing experimental efforts to access the anyonic exchange phase in modified quantum Hall interferometers~\cite{Ehrets.2025,Henzinger.2026}, the setup analyzed here, a Fabry-P\'erot geometry augmented by a single additional QPC, provides a complementary route based on the same co-tunneling physics, and is a promising candidate for the experimental determination of the fractional exchange phase.

\begin{acknowledgments}
This research was supported by the Deutsche Forschungsgemeinschaft (DFG) under Grant No.\@ 406116891 within the Research Training Group RTG 2522/1. Claude Fable 5 and ChatGPT Sol 5.6 were used as research assistants for checking calculations and reviewing the manuscript. This work is based on Chapter 5 of the doctoral thesis \cite{PusterDissertation.2026}, submitted to the Faculty of Physics and Earth System Sciences at Leipzig University in 2026. 
\end{acknowledgments}

\textit{Note added.}---While this manuscript was being finalized, a related analysis of a three-QPC fractional quantum Hall interferometer by Sukhorukov appeared~\cite{Sukhorukov.2026}, with emphasis on testing edge chirality and on extracting the tunneling charge and scaling dimension from the energy dependence of the interference amplitudes; the exchange-phase protocols developed here are complementary to that work.

\appendix
\section{Green functions and evaluation of the Keldysh expression}\label{app:green}

Splitting the expectation value in Eq.~\eqref{eq:IAB_keldysh} into a product of contour-ordered Green functions, each containing only operators from a single edge segment, requires a suitable definition of the contour-ordered Green function. For fermions and bosons, there are universal definitions of suitable contour-ordered Green functions. For anyons, on the other hand, such a definition must be found specifically for each calculation and model. For our interferometer and the quantities we want to compute, a suitable contour-ordered Green function was derived in Ref.~\cite{Kim.2006}. With $\sigma=\pm$ denoting the forward and backward branches, and for operators at $(x,t_{\sigma_1})$ and $(x',t'_{\sigma_2})$, it takes the form
\begin{align}\label{eq:G_contour_app}
    G(x,t_{\sigma_1};x',t'_{\sigma_2})&=e^{i\pi\nu\,\text{sgn}(x-x')\left[\frac{1-\sigma_{12}}{2}\right]}\,\mathcal{G}_{\sigma_{12}}(x-x',t-t')  \nonumber\\
    \mathcal{G}_{\sigma_{12}}(x,t)&=\left[\frac{\frac{k_B T}{\hbar v}\,\epsilon}{2\sin\!\big(\pi \frac{k_B T}{\hbar v}[\epsilon-i\sigma_{12}x+i\sigma_{12}vt]\big)}\right]^\nu ,
\end{align}
where $\sigma_{12}=\frac{1}{2}\left(\sigma_2-\sigma_1+\text{sgn}(t-t')[\sigma_1+\sigma_2]\right)$ encodes the contour ordering, and $\epsilon$ denotes the short-distance cutoff. The Klein factors together with the commutator~\eqref{eq:comm} produce the phase factor in the first line.

Chirality of the edge transport greatly simplifies the branch structure. To see this, consider the retarded Green function,
\begin{align}\label{eq:Gret_app}
    &G^{\rm ret}(x,t)=\Theta(t)\left[G^>(x,t)-G^<(x,t)\right]\nonumber\\
    &\propto \Theta(t)\Bigg[\left(\frac{\tilde{T}_v}{2\sin(\pi \tilde{T}_v[\epsilon-ix+ivt])}\right)^\nu\nonumber\\
    & -e^{i\pi\nu[\text{sgn}(x)-\text{sgn}(x-vt)]}\left(\frac{\tilde{T}_v}{2\sin(\pi \tilde{T}_v[-\epsilon-ix+ivt])}\right)^\nu\Bigg] ,
\end{align}
with $\tilde{T}_v=k_BT/\hbar v$. For $x<0$, one has $\text{sgn}(x)=\text{sgn}(x-vt)=-1$ for all $t>0$, so that the phase factor equals unity and the two terms cancel as $\epsilon\to0$: $G^{\rm ret}(x<0,t)=0$, i.e., forward propagation in time is only possible in the direction of chirality. Similarly, $G^{\rm adv}(x>0,t)=0$. From these two statements, the replacements \eqref{eq:G_t_simplification} and \eqref{eq:G_t_bar_simplification} of the main text follow directly. On the level of the contour-ordered function \eqref{eq:G_contour_app}, they amount to replacing $\sigma_{12}$ by $\tilde{\sigma}_{12}=\frac{1}{2}\left(\sigma_2-\sigma_1+\text{sgn}(x-x')[\sigma_1+\sigma_2]\right)$; we denote the resulting function by $\tilde G$.

Inserting $\tilde G$ into Eq.~\eqref{eq:IAB_keldysh} and extracting the Klein-factor phases, the interference current becomes
\begin{align}\label{eq:IAB_sigma_sum} 
    &I_{3,AB}
    = 
    2\frac{e^*}{\hbar^3}|\gamma|^3
    \,\operatorname{Im}\Bigg[
    \sum_{s_n=\pm}
    \sum_{\sigma_2,\sigma_3=\pm}
    \sigma_2\sigma_3\,e^{i\pi\nu\left[s_n-\frac{\sigma_2+\sigma_3}{2}
    \right]}
    \nonumber\\
    &\times
    \int_{-\infty}^{\infty}dt_2
    \int_{-\infty}^{\infty}dt_3\,e^{i(\tilde V-s_n\Delta\tilde V)t_2}e^{-i(\tilde V+s_n\Delta\tilde V)(t_2-t_3)}
    \nonumber\\
    &\times
    e^{is_n\left(\phi_{AB}+\frac{a}{v}(\tilde{V}_1+\tilde{V}_2+\tilde{V}_3)\right)}
    \,
    \tilde{\mathcal G}(-s_na,0_{-}-t_{3,\sigma_3})
    \nonumber\\
    &\times
    \tilde{\mathcal G}(s_na,0_{+}-t_{2,\sigma_2})
    \,
    \tilde{\mathcal G}(s_na,t_{2,\sigma_2}-t_{3,\sigma_3})
    \Bigg] , 
\end{align}
where $s_n=n-m=\pm1$ labels the two terms of the sum in Eq.~\eqref{eq:IAB_keldysh}, $\tilde{V}=(\tilde{V}_1+\tilde{V}_2)/2 - \tilde{V}_3$, and $\Delta\tilde{V}=(\tilde{V}_1-\tilde{V}_2)/2$. Not all branch configurations contribute: the term with $(s_n,\sigma_2,\sigma_3)=(1,1,-1)$ and the term with $(-1,-1,1)$, after the substitution $t_2\to t_3-t_2$ in the latter, add up to a purely real expression, which drops out of the imaginary part. The same holds for the pair $(1,-1,-1)$ and $(-1,1,1)$. The four surviving assignments, $(s_n,\sigma_2,\sigma_3)=(-1,-1,-1)$, $(-1,+1,-1)$, $(+1,+1,+1)$, and $(+1,-1,+1)$, evaluate the statistical prefactor $\sigma_2\sigma_3\,e^{i\pi\nu\left[s_n-\frac{\sigma_2+\sigma_3}{2}\right]}$ of Eq.~\eqref{eq:IAB_sigma_sum} to $1$, $-e^{-i\pi\nu}$, $1$, and $-e^{i\pi\nu}$, respectively -- precisely the factors carried by the four lines of Eq.~\eqref{eq:IAB_G_products}. 
The four remaining terms can be expressed in terms of the greater Green function,
\begin{widetext}
\begin{align}\label{eq:IAB_G_products} 
    I_{3,AB}
    = 
    2\frac{e^*}{\hbar^3}|\gamma|^3
    \,\operatorname{Im} &
    \int_{-\infty}^{\infty}dt_2 
    \int_{-\infty}^{\infty}dt_3
    \Big[ 
    e^{-i \phi_{AB} }
    e^{i\tilde V_1 \left(t_2-\frac{a}{v}\right)}
    e^{i\tilde V_2 \left(t_3-t_2-\frac{a}{v}\right)}
    e^{i\tilde V_3 \left(-t_3-\frac{a}{v}\right)}
    \,
    \mathcal G^>(-a,t_3)
    \mathcal G^>(a,t_2)
    \mathcal G^>(-a,t_2-t_3)
    \nonumber\\
    &-
    e^{-i \phi_{AB} } 
    e^{-i\pi\nu}
    e^{i\tilde V_1 \left(t_2-\frac{a}{v}\right)}
    e^{i\tilde V_2 \left(t_3-t_2-\frac{a}{v}\right)}
    e^{i\tilde V_3 \left(-t_3-\frac{a}{v}\right)}
    \,
    \mathcal G^>(-a,t_3)
    \mathcal G^>(a,t_2)
    \mathcal G^>(a,t_3-t_2)
    \nonumber\\
    &+
    e^{i \phi_{AB} } 
    e^{i\tilde V_1 \left(t_3-t_2+\frac{a}{v}\right)}
    e^{i\tilde V_2 \left(t_2+\frac{a}{v}\right)}
    e^{i\tilde V_3 \left(-t_3+\frac{a}{v}\right)}
    \,
    \mathcal G^>(-a,-t_3)
    \mathcal G^>(a,-t_2)
    \mathcal G^>(a,t_2-t_3)
    \nonumber\\
    &-
    e^{i \phi_{AB} } 
    e^{i\pi\nu}
    e^{i\tilde V_1 \left(t_3-t_2+\frac{a}{v}\right)}
    e^{i\tilde V_2 \left(t_2+\frac{a}{v}\right)}
    e^{i\tilde V_3 \left(-t_3+\frac{a}{v}\right)}
    \,
    \mathcal G^>(-a,-t_3)
    \mathcal G^>(-a,t_2)
    \mathcal G^>(a,t_2-t_3)
    \Big]. 
\end{align}
\end{widetext}
where $\mathcal{G}^>(x,t)=\mathcal{G}_{\sigma_{12}=1}(x,t)$. Rewriting the four terms as expectation values of tunneling operators yields Eq.~\eqref{eq:IAB_four_terms} of the main text.

\noindent To pass to frequency space, we write
\begin{equation}\label{eq:G_fourier_app}
    \mathcal{G}^>(x,t)=2\left(\frac{\epsilon}{v}\right)^{\!\nu}\int_{-\infty}^{\infty}\frac{d\omega}{2\pi}\, e^{-\epsilon\omega/v}\,e^{i\omega x/v}\,e^{-i\omega t}\,\Omega_r(\omega) ,
\end{equation}
with the real-valued spectral function
\begin{equation}\label{eq:Omega_app}
    \Omega_r(\omega)=\begin{cases}
    \Theta(\omega)\frac{\pi}{(2\pi)^\nu\Gamma(\nu)}|\omega|^{-1+\nu},& T=0 ,\\[0.35cm]
    \frac{\pi \left(\frac{k_B T}{\hbar}\right)^{\nu -1} \frac{e^{\frac{\hbar \omega}{2 k_B T}}}{\cosh \left(\frac{\hbar \omega}{k_B T}\right)-\cos (\pi  \nu )}}{2 \Gamma (\nu ) \Gamma \left(1-\frac{\nu }{2}-\frac{i \hbar \omega}{2 \pi k_B T}\right) \Gamma \left(1-\frac{\nu }{2}+\frac{i \hbar \omega}{2 \pi k_B T}\right)} & T>0 .
    \end{cases}
\end{equation}
The finite-temperature form of $\Omega_r$ is quoted from Ref.~\cite{Puster.2025} (see also Ref.~\cite{Chamon.1995}). In the notation of the main text, $\mathcal{G}^>(\omega)=2\,(\epsilon/v)^{\nu}\,e^{-\epsilon\omega/v}\,\Omega_r(\omega)$, which fixes the prefactor of Eq.~\eqref{eq:G_grt_T0}, and 
combining the spatial Fourier phases of the  oscillator Green functions with the position-dependent zero-mode phases (see Appendix~\ref{app:voltageboson}) gives the common propagation factor $e^{3i a \omega/v  }$ of Eq.~\eqref{eq:IAB_fourier}. 

\section{Treatment of voltages in the bosonization calculation}\label{app:voltageboson}
{
In the main text, the fields $\phi_i$ denote only the oscillatory parts of the chiral boson fields. Here we briefly restore the zero modes and derive the voltage-dependent factors in Eq.~\eqref{eq:Htun}. We follow the finite-size bosonization formalism of Ref.~\cite{geller1997aharonov}, using an auxiliary edge length $L$ that is taken to infinity at the end of the calculation. We denote the complete boson field on edge $i$ by $\Phi_i$ and decompose it as
\begin{align}  
\Phi_i(x,t) = \phi_i(x,t)+\phi_i^{0}(x,t), \label{eq:full_field_decomposition} 
\end{align} 
where $\phi_i$ contains only oscillator modes and satisfies periodic boundary conditions. The zero mode accounts for the total charge and obeys the winding condition 
\begin{align} 
    \phi_i^{0}(x+L,t)-\phi_i^{0}(x,t) = 2\pi N_i . \label{eq:zero_mode_winding} 
\end{align} 
Here 
\begin{align} 
    N_i = \frac{1}{2\pi} \int_0^L dx\,\partial_x\Phi_i(x) 
\end{align} 
measures the edge charge in units of the electron charge $e$. For a Laughlin edge its eigenvalues differ by $\nu$, since a minimal quasiparticle changes the charge by $e^*=\nu e$. For a right-moving edge, the solution of Eq.~\eqref{eq:zero_mode_winding} can be written as~\cite{geller1997aharonov} 
\begin{align} 
    \phi_i^{0}(x) = \frac{2\pi N_i}{L}x-\nu\chi_i,  \label{eq:zero_mode_field} 
\end{align} 
with $[\chi_i,N_j]=i\delta_{ij}$.
Correspondingly, the Hamiltonian separates into an oscillator part and a zero-mode part, 
\begin{align} 
    H_i &= H_{i,\mathrm{osc}}+H_{i,\mathrm{zm}}, \\ H_{i,\mathrm{zm}} &= \frac{\pi\hbar v}{\nu L}N_i^2 . \label{eq:zero_mode_hamiltonian} 
\end{align} 
The sum of the oscillator Hamiltonians $H_0=\sum_i H_{i,\mathrm{osc}}$ is the Hamiltonian  Eq.~\eqref{eq:Heq}. To introduce the source voltages, we assume that the voltage drop occurs at the ideal contacts. The electrostatic reference of the coherent edge region is taken to be fixed, and there is no longitudinal electric field between the source and drain contacts. The source contact therefore fixes the electrochemical potential, and hence the incoming zero-mode charge, while the resulting uniform excess density propagates chirally through the interferometer \cite{law2006electronic}.  The boundary condition imposed by source $i$ is that the edge chemical potential equals $eV_i$. For the uniform zero mode this gives 
\begin{align} 
    \left. \frac{\partial H_{i,\mathrm{zm}}}{\partial N_i} \right|_{N_i=\overline N_i} = eV_i . \label{eq:voltage_boundary_condition} 
\end{align} 
Equivalently, $\overline N_i$ minimizes the grand-canonical zero-mode energy 
\begin{align} 
    K_{i,\mathrm{zm}} = H_{i,\mathrm{zm}}-eV_iN_i = \frac{\pi\hbar v}{\nu L} \left(N_i-\overline N_i\right)^2 +\mathrm{const}. 
\end{align} 
Solving Eq.~\eqref{eq:voltage_boundary_condition} yields \begin{align} 
    \overline N_i &= \frac{\nu eV_iL}{2\pi\hbar v}, \label{eq:mean_zero_mode} 
\end{align} 
and therefore $2\pi\overline N_i/L = \tilde V_i/v$. 
Because the edge is chiral, this incoming density shift is transported unchanged downstream. The zero mode obeys the same chiral equation of motion as the oscillator field, 
\begin{align}   
\left(\partial_t+v\partial_x\right)\phi_i^{0}(x,t)=0. 
\end{align} 
Including a possible non-extensive density fluctuation $\delta N_i$, we can write 
\begin{align} 
    N_i=\overline N_i+\delta N_i  \ \ .
\end{align} 
Using Eq.~\eqref{eq:mean_zero_mode}, the formal solution is
\begin{align} 
    \phi_i^{0}(x,t) = \frac{2\pi\delta N_i}{L}(x-vt) -\nu\chi_i + \frac{\tilde V_i}{v}(x-vt). \label{eq:zero_mode_voltage_solution} 
\end{align} 
An additive constant depending on the choice of coordinate origin has been omitted, which can be absorbed into the  tunneling amplitude.  
The two operator-valued pieces of Eq.~\eqref{eq:zero_mode_voltage_solution} have a distinct role. The factor $e^{-i\nu\chi_i}$ changes $N_i$ by $-\nu$ and therefore annihilates one minimal Laughlin quasiparticle. In the $L\to\infty$ limit it becomes part of the Klein factor $\kappa_i$.  
By contrast, the residual factor proportional to $\delta N_i/L$ vanishes as $L\rightarrow\infty$. The voltage-dependent ratio $\overline N_i/L$, however, remains finite and gives $\phi_i^{0}(x,t)\to \tilde{V}_i\left(\frac{x}{v}-t\right)$.  
Using Eq.~\eqref{eq:zero_mode_voltage_solution}, the complete quasiparticle annihilation operator can be written in the $L\rightarrow\infty$ limit as 
\begin{align} 
    \Psi_i(x,t) &= e^{i \tilde{V}_i\left(\frac{x}{v}-t\right)}\, \psi_i(x,t), \label{eq:full_voltage_qp_operator} \\ 
    \psi_i(x,t) &= (2\pi)^{-\nu/2} \kappa_i e^{i\phi_i(x,t)} . 
\end{align} 
Here $\psi_i$ is  the operator used in the main text. Its Green functions contain only the oscillator modes and are therefore evaluated with the voltage-independent Hamiltonian $H_0$. Substituting Eq.~\eqref{eq:full_voltage_qp_operator} into a tunneling operator gives 
\begin{align}   
&\Psi_m^\dagger(x_{mn},t)\Psi_n(x_{nm},t) \nonumber\\ &\quad = e^{-i(\tilde V_n-\tilde V_m)t} e^{\frac{i}{v} (\tilde V_nx_{nm}-\tilde V_mx_{mn})} \psi_m^\dagger(x_{mn},t)\psi_n(x_{nm},t). \label{eq:dressed_tunneling_operator} 
\end{align} 
The c-number factors may therefore be removed from the quasiparticle fields and placed entirely in the tunneling Hamiltonian, yielding Eq.~\eqref{eq:Htun}. }

\section{Residue summation for the analytic result}\label{app:residues}

Here we sketch the derivation of the closed-form expression \eqref{eq:IAB_analytic} from the frequency integral \eqref{eq:IAB_fourier}. Inserting $\mathcal{G}^>(\omega)=2(\epsilon/v)^{\nu}e^{-\epsilon\omega/v}\Omega_r(\omega)$ with the finite-temperature spectral function \eqref{eq:Omega_app}, each of the four terms of Eq.~\eqref{eq:IAB_fourier} becomes an integral over a product of three $\Omega_r$ functions with shifted arguments, multiplied by the phase factor $e^{3i\omega a/v}$. The cutoff-dependent damping factors combine to $e^{-\epsilon(\omega+\ldots)/v}$. Within the approximation $1-2k_BT\epsilon/\hbar v\approx1$, they are kept only as convergence regulators and dropped otherwise.

The analytic structure of the integrand is simple: the inverse $\Gamma$ functions in Eq.~\eqref{eq:Omega_app} are entire, so the poles of $\Omega_r(\omega)$ at finite temperature are exclusively those of the factor $[\cosh(\hbar\omega/k_BT)-\cos(\pi\nu)]^{-1}$, located on the imaginary axis at
\begin{equation}\label{eq:poles_app}
    \omega^{\pm}_{k}=i\tilde T\left(2\pi k\pm\pi\nu\right),\qquad k\in\mathbb{Z} ,
\end{equation}
with $\tilde T=k_BT/\hbar$. Each of the three factors $\Omega_r(\pm\omega+\ldots)$ therefore contributes one tower of equally spaced poles, shifted by the respective voltage argument. For $a>0$, the factor $e^{3ia\omega/v}$ allows one to close the integration contour in the upper half plane, and the integral equals the sum of the residues at the enclosed pole towers.

Summing the residues along a tower, consecutive poles differ by $2\pi i\tilde T$ in $\omega$, so each tower produces a power series in the variable $e^{3ia(2\pi i \tilde T)}=e^{-6\pi\tilde a\tilde T}$, with $\tilde a=a/v$, precisely the argument of the hypergeometric functions in Eq.~\eqref{eq:IAB_analytic}.  
The series resums into regularized generalized hypergeometric functions $_3\tilde F_2$. Collecting all towers and the four terms of Eq.~\eqref{eq:IAB_fourier} yields the two-term expression \eqref{eq:IAB_analytic}, in which the remaining explicit factors, the trigonometric denominators and the phase factors $e^{i\tilde a(\ldots)}$, originate from the residue of the first pole of each tower and from the voltage shifts of the pole positions. 

As consistency checks, we note the following. The QPC separation $a$ enters Eq.~\eqref{eq:IAB_analytic} only through the thermal attenuation
$e^{-3\pi\nu\tilde a\tilde T}$, the hypergeometric argument $e^{-6\pi\tilde a\tilde T}<1$, which guarantees convergence of the series, and through the explicit phase factors linear in $\tilde a\tilde V_i$, which encode the propagation phases responsible for the non-universal corrections discussed in Sec.~\ref{sec:results}. Additionally, we have validated the result by direct numerical integration of Eq.~\eqref{eq:IAB_fourier}.

\section{Time-of-flight phase corrections in the simultaneous-measurement protocol}
\label{app:tof_simultaneous} 
{For the balanced voltage configuration $V_1=-V_2=V/2$ and $V_3=0$, we consider the phase difference between the two simultaneously measured interference currents  
\begin{equation}
    \delta_2(V)-\delta_3(V)+\pi = \pi\nu+2\delta_{\rm tof}(V) \pmod{2\pi} ,
\end{equation}
where $\delta_{\rm tof}$ is the negative of the phase of the interference amplitude in drain D3. Using the four terms defined in Eq.~\eqref{eq:IAB_fourier}, it can be written as
\begin{align}
\delta_{\rm tof}(V) 
=& 
-\arg  \int_{-\infty}^{\infty}d\omega\,e^{3i\tilde a\omega}
\left(\widetilde I_{\rm I}+\widetilde I_{\rm II}+\widetilde I_{\rm III}+\widetilde I_{\rm IV}\right) .
\label{eq:delta_tof_integral}
\end{align}  
Abbreviating  $x= \tilde V / 4\pi\tilde T$ and  $q=e^{-6\pi\tilde a\tilde T}$,  the residue evaluation of Eq.~\eqref{eq:delta_tof_integral} gives, up to a positive real prefactor,
\begin{equation}
\delta_{\rm tof}(V)
=
-\arg\left\{
e^{i\pi\nu/2}
\left[
A_0(x,q)
-
e^{3i\tilde a\tilde V/2}A_1(x,q)
\right]
\right\},
\label{eq:delta_tof_closed}
\end{equation}
where
\begin{align}
A_0(x,q)
={}&
\frac{
\Gamma(\nu+ix)\Gamma(\nu-ix)
}{
\Gamma(1+ix)\Gamma(1-ix)
}
\,{}_3F_2\!\left(
\begin{matrix}
\nu,\nu+ix,\nu-ix\\
1+ix,1-ix
\end{matrix};q
\right),
\\
A_1(x,q)
={}&
\frac{
\Gamma(\nu-ix)\Gamma(\nu-2ix)
}{
\Gamma(1-ix)\Gamma(1-2ix)
}
\,{}_3F_2\!\left(
\begin{matrix}
\nu,\nu-ix,\nu-2ix\\
1-ix,1-2ix
\end{matrix};q
\right).
\end{align}
The phase in Eq.~\eqref{eq:delta_tof_closed} is followed continuously on the $x>0$ branch. At zero temperature only process I contributes, and Eq.~\eqref{eq:delta_tof_integral} reduces to
\begin{align}
\delta_{\rm tof}(V)
\xrightarrow{T\to0}&
-\arg\int_{0}^{\tilde V/2}d\omega\, 
e^{3i\tilde a\omega}
\mathcal G^>(\omega)
\notag \\ &\times
\mathcal G^>\!\left(\frac{\tilde V}{2}-\omega\right)
\mathcal G^>\!\left(\omega+\frac{\tilde V}{2}\right).
\end{align}
Using $\mathcal G^>(\omega)\propto\Theta(\omega)\omega^{\nu-1}$, rescaling the integration variable, and expanding the propagation factor gives
\begin{equation}
\delta_{\rm tof}(V) = -\frac{3}{2}
\frac{\mathrm{B}\!\left(\frac{\nu+1}{2},\nu\right)}
     {\mathrm{B}\!\left(\frac{\nu}{2},\nu\right)}
\frac{ae^*V}{v\hbar}
+
\mathcal O\!\left[
\left(\frac{ae^*V}{v\hbar}\right)^3
\right],
\end{equation}
which yields the linear voltage dependence used for the extrapolation in the main text.}

\section{Free-fermion benchmark at $\nu=1$}\label{app:nu1}

For $\nu=1$ the edge theory describes free chiral fermions, and the interferometer becomes an elastic single-particle scattering problem. Comparison with the Landauer-B\"uttiker description therefore provides an exact benchmark of the formalism, in particular of the Klein-factor treatment. Using $\Gamma(\frac{1}{2}-iy)\,\Gamma(\frac{1}{2}+iy)=\pi/\cosh(\pi y)$, the spectral function \eqref{eq:Omega_app} reduces at $\nu=1$ to
\begin{equation}\label{eq:Omega_nu1}
    \Omega_r(\omega)\big|_{\nu=1}=\frac{1}{2}\left[1-f(\omega)\right] ,
\end{equation}
with the Fermi function $f(\omega)=[e^{\hbar\omega/k_BT}+1]^{-1}$, so that $\Omega_r(\omega)+\Omega_r(-\omega)=1/2$. Since $-e^{i\pi\nu}\to+1$ at $\nu=1$, the co-tunneling terms in Eq.~\eqref{eq:IAB_fourier} enter with the same sign as their direct partners, and, within the cutoff approximation of Appendix~\ref{app:residues}, the pairs $(\tilde I_{\rm I},\tilde I_{\rm II})$ and $(\tilde I_{\rm III},\tilde I_{\rm IV})$ combine through $\mathcal{G}^>(\omega-\tilde V_2)+\mathcal{G}^>(\tilde V_2-\omega)=\epsilon/v$. Therefore, the spectral weight of the interference current becomes independent of $V_2$. Using furthermore $\Omega_r(\omega-\tilde{V}_3)\,\Omega_r(\tilde V_1-\omega)-\Omega_r(\tilde{V}_3-\omega)\,\Omega_r(\omega-\tilde V_1)=\frac{1}{4}\big[f(\omega-\tilde V_1)-f(\omega-\tilde{V}_3)\big]$, the four terms collapse to
\begin{align}\label{eq:IAB_nu1}
   I_{3,AB} \big|_{\nu=1}= &-2e^*\lambda^3\!\int_{-\infty}^{\infty}\!\frac{d\omega}{2\pi}\,\big[f(\omega-\tilde V_1)-f(\omega-\tilde{V}_3)\big]\nonumber\\
    &\times\,\text{Im}\,e^{i\phi_{AB}+3i\frac{a}{v} \omega } ,
\end{align}
with the dimensionless tunneling amplitude $\lambda=|\gamma|\epsilon/\hbar v$.

Equation~\eqref{eq:IAB_nu1} is the Landauer-B\"uttiker interference current of free chiral fermions. An electron incident from source S1 reaches drain D3 along two paths which enclose the AB flux with a path-length difference of $3a$. Either directly, by tunneling at the QPC connecting segments 1 and 3, or via segment 2, by tunneling at the two remaining QPCs.   
An electron incident from source S2, by contrast, passes the QPC connecting segments 1 and 2 before the one connecting segments 2 and 3. If it tunnels into segment 1, it flows into drain D1 without a second opportunity to reach segment 3. The geometry thus admits no interfering path from S2, and consistently the occupation factor in Eq.~\eqref{eq:IAB_nu1} involves only $V_1$ and $V_3$.  We finally note that taking the limit $\nu\to1$ in the closed form \eqref{eq:IAB_analytic} is not straightforward. 
Pairs of poles in Eq.~\eqref{eq:poles_app}, $\omega^{+}_{k}$ and $\omega^{-}_{k+1}$, merge at $\nu=1$, and the limit is therefore better  
performed starting from Eq.~\eqref{eq:IAB_fourier} rather than Eq.~\eqref{eq:IAB_analytic}.

%

\end{document}